\documentclass[preprint2]{aastex}

\usepackage{graphicx}
\usepackage{txfonts}
\usepackage{cite}
\usepackage{times}
\usepackage{natbib}

\slugcomment{Not to appear in Nonlearned J., 45.}

\shorttitle{Global modelling of X-ray spectra produced in O-type star winds}
\shortauthors{Herv\'e et al.}

\begin{document}


        \title{Global modelling of X-ray spectra produced in O-type star winds}
		\author{A.\ Herv\'e \and G.\ Rauw  \and Y.\ Naz\'e\thanks{Research Associate FRS-FNRS (Belgium)}}
		\affil{GAPHE, D\'epartement AGO, Universit\'e de Li\`ege, All\'ee du 6 Ao\^ut 17, B\^at. B5c, 4000 Li\`ege, Belgium}
              \email{herve@astro.ulg.ac.be} 
              
              \and
              
     	\author{A. Foster}         
    	\affil{Smithsonian Astrophysical Observatory, 60 Garden Street, Cambridge MA 02138, USA} 
		\email{afoster@cfa.harvard.edu } 




\begin{abstract}
High-resolution X-ray spectra of O-type stars revealed less wind absorption than expected from smooth winds with conventional mass-loss rates. Various solutions have been proposed, including porous winds, optically thick clumps or an overall reduction of the mass-loss rates. The latter has a strong impact on the evolution of the star. Our final goal is to analyse high resolution X-ray spectra of O-type stars with a multi temperature plasma model in order to determine crucial stellar and wind parameters such as the mass loss rate, the CNO abundances and the X-ray temperature plasma distribution in the wind. In this context we are developing a modelling tool to calculate synthetic X-ray spectra. We present, here, the main ingredients and physics necessary for a such work.
Our code uses the most recent version of the AtomDB emissivities to compute the intrinsic emissivity of the hot plasma as well as the CMFGEN model atmosphere code to evaluate the opacity of the cool wind. Following the comparison between two formalisms of stellar wind fragmentation, we introduce, for the first time in X-rays, the effects of a tenuous inter-clump medium. We then explore the quantitative impact of different model parameters on the X-ray spectra such as the position in the wind of the X-ray emitting plasma. For the first time, we also show that the two formalisms of stellar wind fragmentation yield different results, although the differences for individual lines are small and can probably not be tested with the current generation of X-ray telescopes. 
As an illustration of our method, we compare various synthetic line profiles to the observed
 \ion{O}{8} $\lambda$ 18.97 $\AA$ line in the spectrum of $\zeta$\,Puppis.
We illustrate how different combinations of parameters can actually lead to the same morphology of a single line, underlining the need to analyse the whole spectrum in a consistent way when attempting to constrain the parameters of the wind.

\end{abstract}





\section{Introduction}

The study of stellar winds is crucial for our understanding of the evolution of massive stars and their interactions with their environment. In fact, mass-loss plays a fundamental role in the evolution of massive stars (e.g.\ \citealt{Hirschi}) as well as in the mechanical and chemical feedback of these stars into the interstellar medium (e.g.\ \citealt{Dopita}). Over the last decade, our view on the mass-loss of massive stars has changed as the importance of wind fragmentation became clear.

Direct observational evidence for the existence of clumps in the winds of O-type stars was obtained from the low-level, rapid variability of the \ion{He}{2} $\lambda$\,4686 emission line in the spectrum of $\zeta$\,Puppis (\citealt{ELM}). Indirect evidence stems from the analysis of the spectra of O-type stars. Indeed, with observations from the $FUSE$ satellite, it became clear that the modelling of lines of ions such as \ion{O}{5}, \ion{N}{4} and \ion{P}{5} with model atmosphere codes such as CMFGEN, is impossible if clumping is not included (e.g. \citealt{BLH}).

The existence of such small-scale wind structures has important consequences on many diagnostics of mass-loss rates. Indeed, strong clumping often implies a substantial reduction, by factors between a few and about ten, of the overall mass-loss rate compared to a homogeneous stellar wind. 
In this context, our ignorance of the properties of these wind structures translates into uncomfortably large uncertainties on the actual mass-loss rates. For instance, whilst the space between clumps is generally supposed to be totally empty, some recent studies \citep{Zsargo,Sundqvist} showed that it is necessary to include the effects of a tenuous inter-clump medium to correctly predict the strength of various lines in the observed UV spectra of O-type stars. By doing so, these authors even found that the observed profiles might again be compatible with the conventional mass-loss rates.

Mass-loss rate diagnostics in different wavelength domains are strongly sensitive to different aspects of the wind fragmentation. For instance, H$_{\alpha}$ and radio emission are collisional processes which scale with density squared and are thus sensitive to the clumping factor regardless of the size of the clumps. In clumped winds with a void inter-clump medium, the velocity law is intermittent and some velocities are not present along a given sight line. This porosity in the velocity field is called 'vorosity' \citep{owockiliege}. The latter strongly modifies the Sobolev length and its impact is mainly visible on the UV resonance lines.

The present paper deals with the impact of wind structure on the morphology of X-ray spectra of massive stars. High-resolution $Chandra$ and $XMM-Newton$ X-ray spectra of presumably single O-type stars revealed less absorption of the X-rays by the cool wind than expected for a homogeneous wind with conventional mass-loss rates. This property can be explained in the case of fragmented stellar winds (e.g. \citealt{Miller}; \citealt{KCO}; \citealt{OFH}; see also \citealt{GN} for a review), although  \citet{Cohen} argue in favour of an overall reduction of the mass-loss rate of the wind without the need to include porosity effects.

In a fragmented wind with optically thick clumps, X-ray lines are sensitive to porosity, which depends on clump continuum optical depth, and thus clump size as well as the clumping factor. That is why high-resolution X-ray spectra are important tools to study the structure of the wind and help us determine mass loss rates which must be consistent with the other wavelength domain analyses. Previous works in the X-ray domain (e.g. \citealt{OFH}), were performed on a line-by-line basis; the spectra were never analysed as a whole at once. Using multi-temperature plasma models in future work, the global fit of observations will give information concerning the different shock temperatures as well as the size and the position of the different X-ray emitting shells.

With this in mind, we have designed a new code for the purpose of simulating the full high-resolution X-ray spectrum of a single massive star. Our method properly accounts for the effect of line blending on the line profiles, and allows testing the consistency of current models for the X-ray emission of massive stars over the full range of the X-ray domain. Therefore, our code is meant as a further step towards extracting the maximum amount of information from the high-resolution X-ray spectra. In this paper, we introduce the different free parameters and investigate their impact on single temperature synthetic X-ray spectra. We further discuss the main assumptions and physics in our code and compare different formulations of the wind fragmentation. In particular, we investigate for the first time the impact of a tenuous, but non-void, inter-clump medium on the synthetic X-ray spectra. Finally, we present the first application of our code to fit the region around the \ion{O}{8} $\lambda$ 18.97$\AA$ line in the $XMM$ spectrum of $\zeta$\,Puppis and illustrate the degeneracies on the stellar wind parameters that exist when a single line is fitted. 
In future work, we will apply this code to the full X-ray band spectra of $\zeta$\,Puppis. 

\section{Assumptions and main concepts of the code}

Before using a multi temperature plasma model to analyse the X-ray spectra of O-type stars in detail, we need to calculate single temperature synthetic spectra with the most recent stellar wind theories (emission and absorption) and with as few as possible free parameters. We decide to follow the wind embedded shocks scenario. Consequently, we need to know where the shocks arise in the wind and what is the temperature of the resulting post-shock hot plasma. Using up-to-date emissivities, we can then determine the X-ray emission of a hot plasma as a function of temperature, abundances and quantity of matter present in the shocks. Next, we determine how the cool wind material absorbs the X-ray photons and how the wind structure (i.e homogeneous or porous) impacts on the emergent X-ray flux. In the following subsections, we describe the tools and assumptions considered to build our code and to calculate synthetic spectra.

\subsection{The plasma emission spectrum}

Previous works on X-ray line profile fitting have always normalized the line profiles, i.e the overall line flux was never physically modelled. We attempt a different approach here, by using the true X-ray emissivity of an optically thin plasma. For this purpose, we use AtomDB\footnote{ This was previously known as the APEC model \citep{smith2001}. The interested reader is invited to visit the section 'physics' on the AtomDB website for a precise description of the calculation of the continuum and lines in a hot plasma  at the url http://www.atomdb.org/} which models the emission from a collisionally ionized, optically thin isothermal plasma in thermal equilibrium. Many processes are included in the model: collisional excitation, ionization, and dielectronic recombination drive the ionization balance and emission spectra from the plasma. Continuum processes such as two-photon decay and bremsstrahlung are also included. Photon-driven processes such as photo-excitation and photo-ionization are not included in the model. The assumption of an optically thin plasma requires a low density ($\le 10^{13}$cm$^{-3}$), therefore three body effects are not included.
The standard AtomDB model data has been recalculated on a finer 0.025 keV/step electron temperature grid for this analysis and we use an energy resolution of 0.2 eV.

\begin{figure}[p]
\includegraphics[angle=90,width=9.cm]{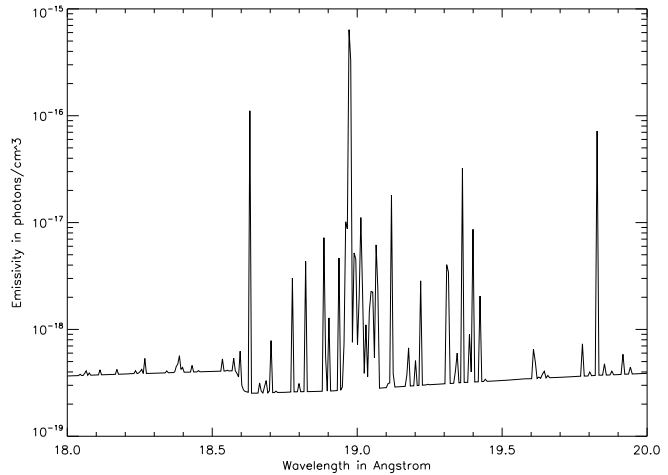}
\caption{Theoretical emission of a plasma with a temperature of kT=0.273 keV and with a $\frac{O_{*}}{O_{\sun}} = 0.3$ and $\frac{N_{*}}{N_{\sun}} = 4.$ by number. \label{figpre2}}
\end{figure}

In addition to the strongest lines that dominate the spectrum, the high-resolution theoretical spectra reveal the presence of numerous weak lines and also discontinuities in the continuum (Fig. \ref{figpre2}). Their contribution to the emergent flux is non-null and it is necessary to take them into account for the computation of emergent spectra as they certainly modify the emissivities and profiles of the strongest observed lines. \linebreak
When fitting actual observations, one further needs to account for the presence of plasma at different temperatures, located at different places in the wind. Therefore the normalization of the model will result from the combination of several model spectra weighted according to their relative emission measure. This will be dealt with in our forthcoming paper on model fits of $\zeta$\,Puppis spectrum.\linebreak
In addition, in the X-ray emitting plasma, some ions are present in a helium-like form. The treatment of the transition between the $2\,^{3}S_{1}$ and $2\,^{3}P_{j}$ levels must be done separately. Indeed, the presence of a strong UV radiation modifies the populations of the metastable upper level of the forbidden line via a pumping to the upper level of the inter-combination line (\citealt{blue71}; \citealt{porquet01}). 
We will discuss this fundamental effect in our more complete analysis of $\zeta$\,Puppis presented in a forthcoming paper (Herv\'e et al., in preparation).

We consider the hot plasma to be embedded in the cool wind and to move outwards along with the latter. We decompose the emitting volume into small cells where we calculate the X-ray emission of each cell and its shift in wavelength in the observer's frame of reference due to the overall wind motion.

\subsection{Wind optical depth and mass-absorption coefficient}

The X-ray emitting plasma is assumed to be optically thin to X-rays. An X-ray photon emitted by the hot plasma, either escapes from the wind or is absorbed by the cool wind material. Thus the radiative transfer is reduced to the calculation of $\tau$, the optical depth along the line-of-sight due to the cool wind material. Following the approach of \citet{OC01}, the absorption by the stellar wind  is essentially characterized by the $\tau_{*}$ parameter and the mass-absorption coefficient $\kappa$, which are linked by the relation $\tau_{*}\equiv \frac{\kappa \dot{M}}{4\pi v_{\infty}R_{\ast}}$. These absorption coefficients are dependent on the wavelength and the abundances of the different elements. To evaluate the mass-absorption coefficients, we use the radiative transfer code CMFGEN (Hillier $\&$ Miller 1998) which computes the ionization structure of the cool wind  as a function of temperature, abundances of many elements and mass-loss rate. In the ionization calculations, CMFGEN accounts for  different physical processes such as collisional excitation,  photo-ionisation by photospheric  light and X-rays,  and radiative and dielectronic recombination. From the known ionization structure, we can then compute $\kappa$. In the X-ray domain, the opacity is dominated by the photo-ionization of the metals. This absorption is by far the dominant source of opacity and we can safely neglect other physical processes such as free electron scattering. 

\begin{figure}[p]
\includegraphics[angle=90,width=8.cm]{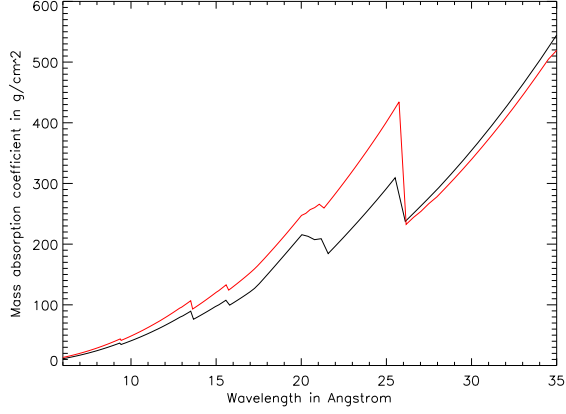}
\caption{Evolution of the mass-absorption coefficient as a function of the wavelength. In black, the mass absorption coefficient with our preliminary results on the abundances determined in the X-ray spectrum of $\zeta$\,Puppis reveals a less opaque wind than with the abundances determined in the most recent UV/optical analysis (in red, J-C Bouret, private communication, Bouret et al., in preparation).\label{figpre1}}

\end{figure}

The wavelength dependence of $\kappa$ (Fig. \ref{figpre1}) and therefore $\tau_*$, is quite sensitive to the model elemental abundaces and to the ionization fraction of singly ionized helium. In this respect, we stress that the large values of the mass absorption coefficient, used in the present paper, reflect the fact that the sum of the CNO abundances in the best-fit model of the UV and optical spectrum of $\zeta$\,Puppis is supersolar (Bouret et al., in preparation).   

The detailed behaviour of $\kappa$ as a function of radius and wavelength (especially at longer wavelengths) also depends upon the ionization structure of the wind and more specifically on the details of the recombination of He$^{2+}$ into He$^+$ \citep{Leutenegger10}. However, for simplicity, we assume here that $\kappa$ is independent of the position in the wind\footnote{This assumption allows us to take $\kappa$ out of the optical depth integral (Eq.\,\ref{eqn3}) and thus to achieve an analytical evaluation of this integral.} and we evaluate its value at $r = 10$\,R$_{*}$. In our detailed analysis of the RGS spectrum of $\zeta$\,Puppis (Herv\'e et al., in preparation), we will drop this assumption and rather introduce a radial dependence of $\kappa$.

Finally, the values of the mass-absorption coefficient or the wind optical depth as a function of the wavelength are used in Eq.\,\ref{eqn3} and \ref{eqn4} (see Sect.\,\ref{fragment} below) to determine the absorption of X-rays by the stellar wind.

\subsection{X-ray absorption by a fragmented wind \label{fragment}}

Previous studies of stellar winds revealed that they are not homogeneous but fragmented. The absorption of X-rays by a clumpy wind has been addressed by several authors who adopted slightly different formalisms for the same phenomenon. The \textit{porosity length prescription} is based on the concept of the mean free path of photons between two successive interactions with clumps, and the opacity of the latter \citep{OC06}. In the \textit{fragmentation frequency prescription}, the opacity of the wind in the X-ray band is estimated by the frequency of clumps passing through a reference radius \citep{OFH}. In this section, we briefly summarize the main aspects of these models.


At the microscopic level, the absorption of X-rays by a stellar wind is mainly due to the bound-free transitions from the K and L energy levels. To start, we assume that the clumps are spherical and very small. We consider clumps over a wide range of optical depth, from optically thin to optically thick. 
In a situation where the individual clumps are optically thick, the effective opacity is essentially determined by the geometrical properties of the clumps \citep{Feld03} and can be written  $\kappa_{eff}=\frac{l^{2}}{m_{c}}=\frac{\kappa}{\tau_{c}} $ where $\kappa$ is the mass-absorption coefficient, $l$ is the geometrical scale of the clump, $m_{c}$ its mass and $\tau_{c}$ its optical depth. This latter can be written as \citep{OC06}:
\begin{equation}
 \tau _{c} =\kappa \langle\rho\rangle \frac{l}{f}=\frac{\kappa \dot{M}}{4\pi r^{2}}\frac{1}{v}\frac{l}{f} 
\end{equation}

where $\langle\rho\rangle$ is the mean density of the stellar wind, $f$ the volume filling factor, $\dot{M}$ is the mass-loss rate of the star and $v$ the velocity of the wind at the radius $r$ which is determined by a  $\beta$-law (i.e $v(r)=v_{\infty}(1-\frac{R_{*}}{r})^{\beta}$ with $\beta=1$ throughout this work).
\\ 
Under this assumption, it appears that the effective reduction depends on the ratio between the clump scale and filling factor, also called the porosity length \citep{OC06}.
The concept of effective opacity can be extended to the optically thin limit using a simple bridging law \citep{OC06}: 
\begin{equation}
\frac{\kappa_{eff}}{\kappa}=\frac{1}{1+\tau_{c}} 
\label{eqn2}
\end{equation}
Using this result in a steady state (i.e.\ non variable) wind, one obtains the wind optical depth in conventional $(p,z)$ coordinates \citep{OC06}:

\begin{equation}
\tau(p,z)=\tau_{\ast} \int^{\infty}_{z} \frac{R_{\ast}dz'}{r'(r'-R_{\ast})+\tau_{\ast}h(r')}
\label{eqn3}
\end{equation}
where $r'=\sqrt{p^{2}+z'^{2}}$, $\tau_{\ast}\equiv \frac{\kappa \dot{M}}{4\pi v_{\infty}R_{\ast}}$ is the optical depth parameter of the wind, $v_{\infty}$ is the terminal velocity of the wind, $R_{*}$ is the radius of the star and $h(r)\equiv\frac{l}{f}$ the porosity length. In the following we adopt $h(r)=h\times r$.


The fragmentation of stellar wind shells is a consequence of hydrodynamical instabilities. Rather than being spherically symmetric, the resulting clumps could actually have a flattened, elongated shape. Using the number $n_{0}$ of fragments passing through some reference radius per unit time, also called the \textit{fragmentation frequency}, and the opacity of each fragment at this reference radius, \citet{OFH} showed that the wind opacity can be written in the general case as:
\begin{equation}
 \tau (p,z)=n_{0}\int^{z_{max}}_{z} (1-e^{-\tau_{j}}) |\mu (r')|\frac{dz'}{v(r')}
\label{eqn4}
\end{equation}
where  $\tau_{j}$ is the optical depth of a flattened fragment $j$ along the line-of-sight and $\mu(r')=\frac{z'}{\sqrt{z'^{2}+p^{2}}}$ represents its orientation. The optical depth along the line-of-sight:
\begin{equation}
 \tau_{j} = \frac{\tau_{j}^{rad}}{|\mu|} = \frac{\kappa \dot{M}}{4\pi}\frac{1}{r^{2}}\frac{1}{n_{0}}\frac{1}{|\mu|}
\label{eqn5}
\end{equation}
with $\kappa$ the mass-absorption coefficient and $\tau_{j}^{rad}$ the average radial optical depth of a fragment located at distance $r$ \citep{OFH04}.
In the case of an isotropic opacity (i.e.\ small-scale spherically symmetric clumps), $\mu$ is taken equal to unity (i.e. the line-of-sight crosses each clump along a radial direction).

In Section 3.1, we compare the profiles obtained with $\tau$ given by Eq.\,\ref{eqn4} to those computed with Eq.\,\ref{eqn3}, first in the case of an isotropic opacity where $\mu$ is taken equal to one, then in a case of  an anisotropic opacity where the clumps are assumed to be flattened (i.e $\mu \neq 1$).

\subsubsection{Differences and common features of the different prescriptions}

As we will show below, the porosity length prescription and the fragmentation frequency prescription are essentially equivalent to first order, although in their original work, \citet{OC06} concentrated on spherical clumps, whilst \citet{OFH} accounted for flattened structures by introducing the line-of-sight orientation via the $\mu$ parameter\footnote{Note that this parameter could also easily be inserted into Eq.\,\ref{eqn2}.}. Actually, the geometry assumed by \citet{OC06} is even more idealized than a pure spherical geometry since it assumes that the line-of-sight always crosses the clump along the radial direction. Strictly speaking, this is valid only if the clumps are very small. The main differences between the two approaches hence concern the geometry of the wind fragments and the way the optical depth of a clump is included in the model, i.e.\ Eq.\,\ref{eqn2} for the porosity prescription, and the $(1-e^{-\tau_{j}})$ term in Eq.\,\ref{eqn4} for the fragmentation frequency. For small optical depths and isotropic clumps ($\mu = 1$), both models should be essentially equivalent.

In a first step, we reproduce the profile of various lines (Table\,\ref{tab1}) for each prescription and compare them with each other and with the results for a homogeneous wind model.
We consider a distribution of spherically symmetric clumps and we thus set $\mu = 1$ in Eq.\,\ref{eqn4} and \ref{eqn5}. Our general results agree well with those of previous studies. First, the larger the porosity of the wind, the larger the mean free path of a photon in the stellar wind and the lower the absorption  \citep{OC06}. Second, the larger the fragmentation frequency, the lower the opacity in the fragmentation prescription, see \citet{OFH}.
 
We have simulated four specific lines with $n_{0}=1.7$ $10^{-4}$\,s$^{-1}$ adopted from \citet{OFH}. This parameter is found by the authors to best fit the \ion{O}{8} $\lambda$ 18.97 $\AA$ line in the HETG and RGS spectra of $\zeta$\,Puppis. Then we derive the porosity length, $h$, needed to reproduce the line profiles computed with this fragmentation frequency (Fig.\ \ref{fig1}). For different values of $\tau_{*}$, hence different values of wavelength (see Table 1), the two line profiles are, in each case, very close.
Note however that we obtain different values for the porosity length when matching different lines computed with the same fragmentation frequency: we have to decrease the porosity parameter when $\tau_{\star}$ increases. In other words, when we compare simulated global spectra using a single porosity parameter to those obtained for a specific value of the fragmentation frequency, we find that the spectra are exactly the same in the wavelength band near the line used to find the parameter $h$ that best matches the value of $n_0$ (in our case the \ion{O}{8} $\lambda$ 18.9 $\AA$ line), but differ over the other wavelength domains (Fig.\,\ref{fig2}). More specifically, the porosity model predicts a slightly lower flux in the short wavelength domain and a higher flux in the long wavelength band.

There are some underlying assumptions in both models, and with current data it seems difficult to tell which prescription provides the better description of reality. The porosity prescription has the advantage to provide an analytical form of the opacity that speeds up the computation. Therefore, in the following, we concentrate mainly on the porosity prescription. The overall results and conclusions obtained in the study for the different parameters, in the next section, are the same for the fragmentation model in an isotropic ($\mu =1$) case.

\section{The impact of model parameters on the emergent spectrum}

For each simulation in this section, we use a terminal velocity $v_{\infty}$ of 2250\,km\,s$^{-1}$. We further assume a $kT_{plasma}$ of 0.217\,keV, solar abundances for the chemical composition of the plasma, a stellar radius $R_{*}$ of 18.6\,$R_{\sun}$ and a mass-loss rate of 4.2 $10^{-6}\,M_{\sun}$\,yr$^{-1}$. These stellar and wind parameters correspond to those used by Oskinova et al (2006) in their work to fit lines of $\zeta$\,Puppis, except for the plasma temperature which corresponds to the closest value in AtomDB to the result of \citet{Hillier93}.

In the remainder of this section, we simulate a series of spectra for such a single O star in order to illustrate the sensitivity of the synthetic spectra on the different model parameters.

\subsection{Porosity}
\begin{figure}[p]
\includegraphics[angle=90,width=8.cm]{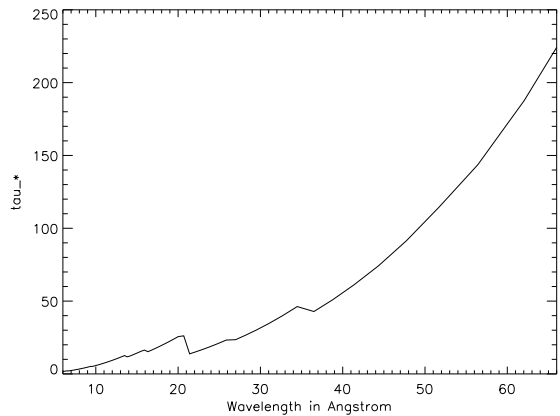}
\caption{$\tau_{*}$ computed with solar abundances, $\dot{M}$ = 4.2 10$^{-6}$ M$_{\sun}$yr$^{-1}$, v$_{\infty}$ = 2250\,km\,s$^{-1}$ and R$_{*}$ = 18.6 R$_{\sun}$. These values are used in Section 3.
\label{tau0}}
\end{figure}

\begin{figure}[p]
\includegraphics[angle=90,width=9.cm]{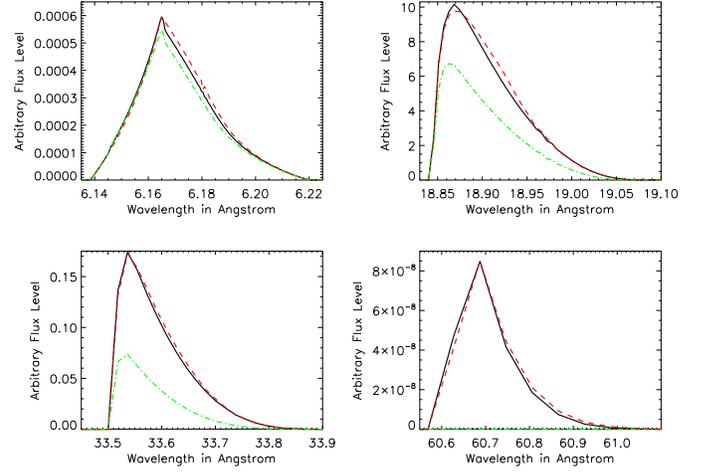}
\caption{Illustration of the impact of porosity ($h=0.07$, 0.07, 0.055 and 0.043\,$R_*$, dashed red line, from left to right and top to bottom) on the profiles of four specific lines (see Table 1). The profiles are compared with those computed in the fragmentation prescription for $n_{0} = 1.7 10^{-4}$\,s$^{-1}$ (black solid line) and for a homogeneous wind model (dot-dashed green line). \label{fig1}}
\end{figure}

\begin{figure}
 \includegraphics[angle=90,width=9.cm]{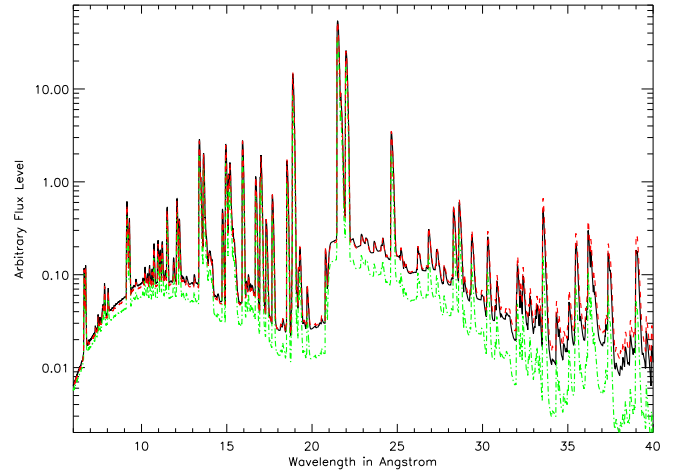}
\caption{Comparison between a synthetic spectrum computed with the porosity prescription adopting $h = 0.07$\,$R_*$ (dashed red line), a spectrum computed with the fragmentation frequency theory with $n_0 = 1.7 10^{-4}$\,s$^{-1}$ (black solid line) and a homogeneous wind (dot-dashed green line). \label{fig2}}
\end{figure}

\begin{figure}
 \includegraphics[angle=90,width=9cm]{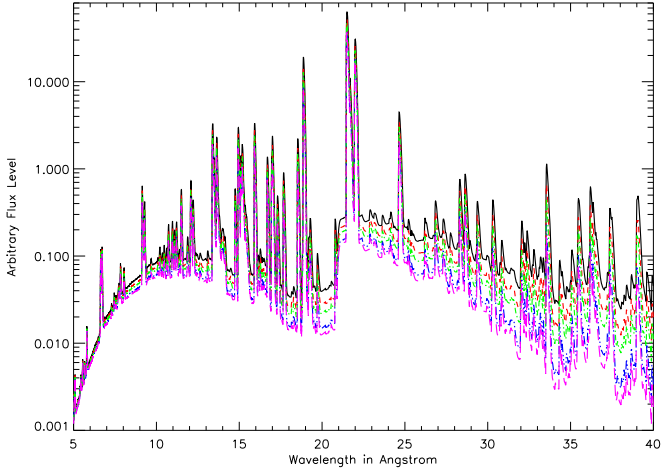}
\caption{Effect of porosity on the emergent spectrum. In long-dashed magenta line, the results for a homogeneous wind, in dashed-double-dot blue line, dashed-dot green line, dashed red line and solid black line the spectra for different porosity parameters (0.02, 0.07, 0.1 and 0.15\,$R_*$, respectively). A porosity of 0.15\,$R_*$ corresponds to the brightest model, whilst 0.02\,$R_*$ corresponds to the most absorbed (hence weakest) model. See the online version of the article for a colour figure.\label{fig5}}
\end{figure}

The first explored parameter is the porosity length. The impact of this parameter for broad band spectra was never presented, in the literature. Fig.\,\ref{fig5} illustrates our results. We have chosen to explore a wide range of values of the parameter $h$. As expected, we find a strong impact on the emergent flux. At the shortest wavelengths, where $\kappa$ is lowest, the porosity length has little impact on the spectrum. However, the situation is radically different at longer wavelengths, where the wind opacity $\kappa$ becomes huge (Fig.\ref{tau0}, Tab.\ref{tab1}). Here, the impact of the porosity length is much more important. We find differences of several orders of magnitude between the spectrum for a homogeneous wind and the simulations with the largest porosity. For a highly porous wind, the variations of $h$ have less impact as the wind is already so fragmented that the photons have few remaining interactions with the clumps.

\begin{figure}
 \includegraphics[angle=90,width=9cm]{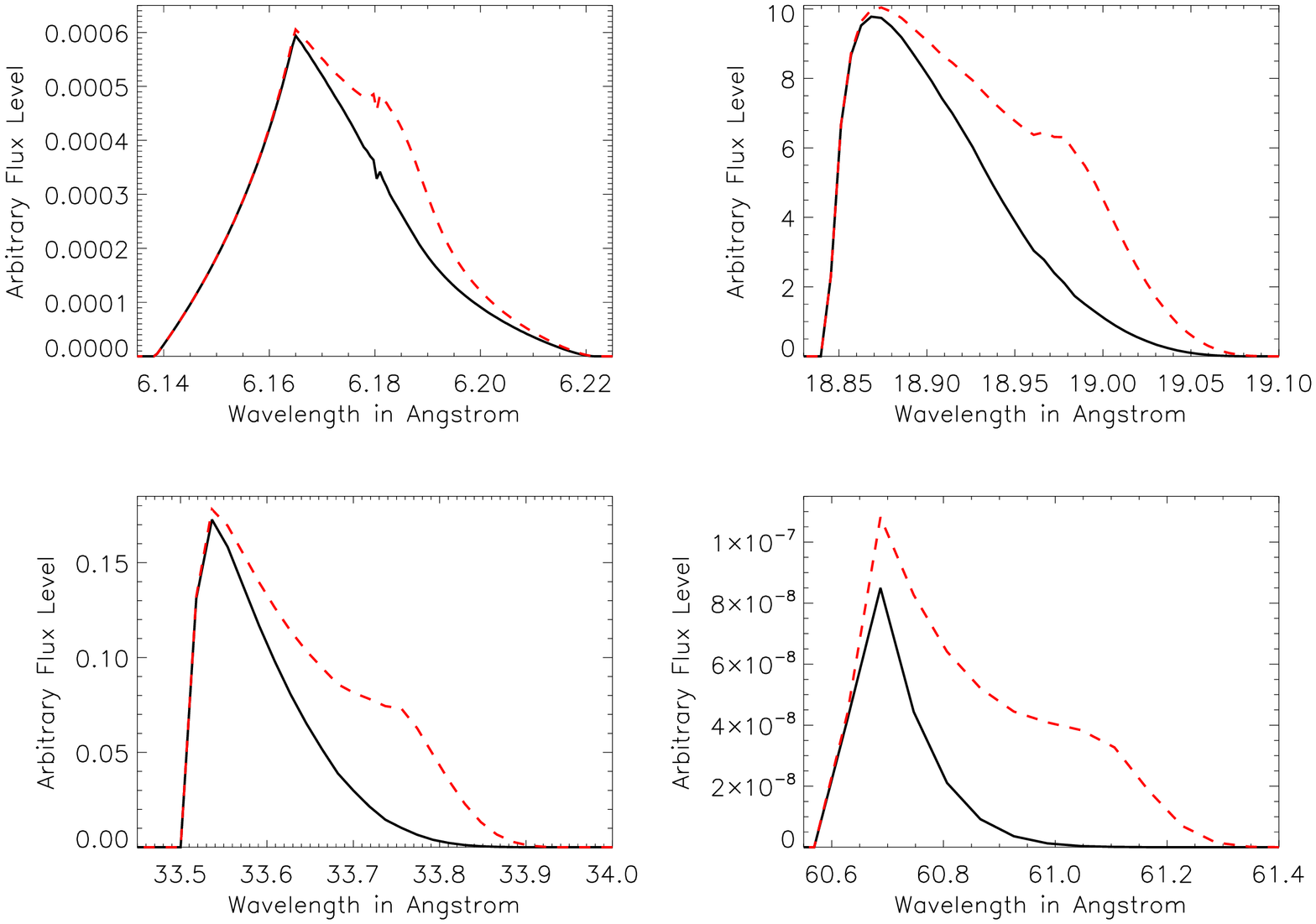}
\caption{Impact of the anisotropy of the opacity (fragmented shells seen under a direction cosine $\mu$) on the emergent line profile, shown as the dashed red line, for specific lines (see Table 1). The results are compared to the line profiles computed for an isotropic opacity ($\mu=1$; solid black line) with $n_0 = 1.7 10^{-4}$\,s$^{-1}$.
\label{fig3}}
\end{figure}

\begin{figure}
 \includegraphics[angle=90,width=9.cm]{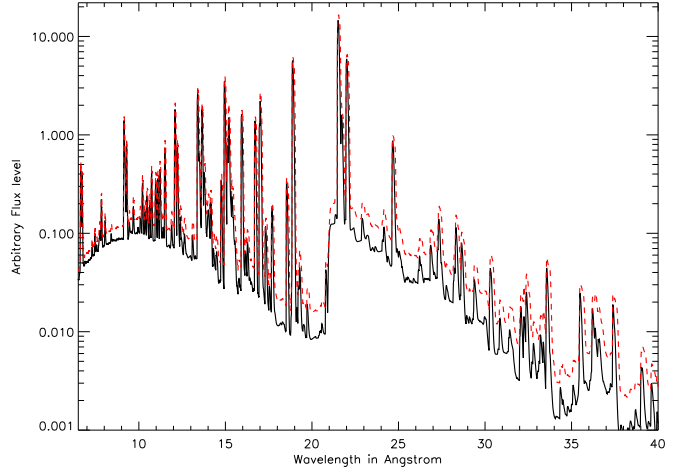}
\caption{Same as Fig.\,\ref{fig3}, but this time for the entire X-ray spectrum. 
\label{fig4}}
\end{figure}

As a second step, we consider the effect of the shape of fragmented shells. For this purpose, the factor $\mu$ in Eq.\,\ref{eqn4} and \ref{eqn5}, which represents the orientation of the fragmented shells with respect to the line-of-sight, is re-introduced into our simulations. Compared to the isotropic case, the opacity of the wind decreases (see Fig.\,\ref{fig3}). The radiation from the back side of the star is less attenuated than in the case of isotropic clumps. Indeed, the fact that fragments are flattened and that photons coming from the rear side of the wind cross these shells mainly perpendicularly implies that they have to cross less matter, and are thus less absorbed, than in the isotropic case. As a result, the total flux and intensity are higher in the anisotropic case. Also, the profiles are less asymmetrical and the full width at half maximum increases. The consequences are more intense spectra  (by a factor 1.2 in the short wavelengths and 3.0 in the longer wavelengths domain) and an increased overlap of lines in the anisotropic model (see Fig.\,\ref{fig4}).

\subsection{Outer radius of the emitting region}
\begin{figure}
 \includegraphics[width=9cm]{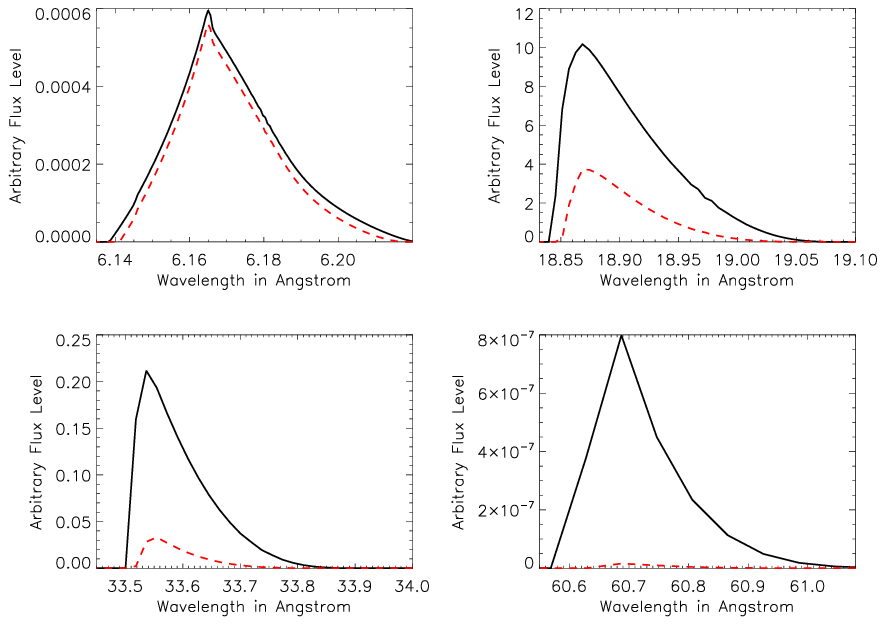}
\caption{Impact of the size of the emitting plasma on the line profiles at different wavelengths. The solid black (respectively dashed red) profiles correspond to an outer radius of the emitting region of 10$R_{*}$ (resp.\ 5$R_{*}$).\label{fig6}}
\end{figure}

\begin{figure}
 \includegraphics[angle=90,width=9cm]{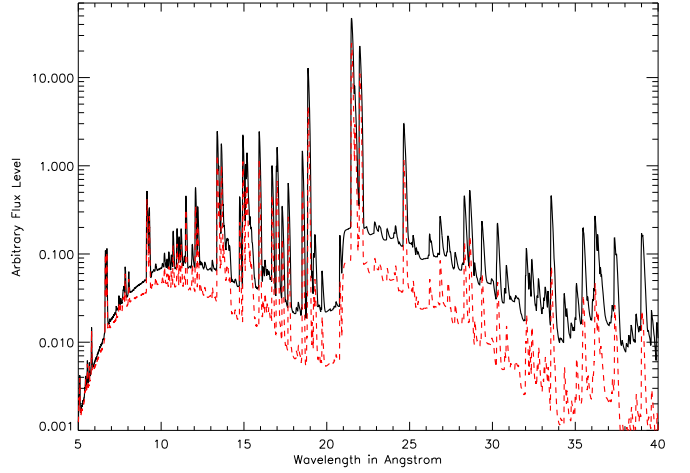}
   \caption{Impact of the size of the emitting plasma on the overall X-ray spectrum. The colours have the same meaning as in Fig.\,\ref{fig6}.\label{fig7}}
\end{figure}

Another parameter of our model is the size of the X-ray emitting plasma ($R_{out}$). As the emissivity is proportional to $\rho^{2}$, the outer parts of the wind contribute less to the total intrinsic (i.e.\ before wind absorption is accounted for) X-ray emission, but cannot be neglected for the wind-absorbed emission as can be seen for specific line profiles in Fig.\,\ref{fig6}. The effect is by far strongest in the long wavelength domain where $\kappa_{\lambda}$ is largest. The intensity of the lines in Fig.\,\ref{fig6} varies by a factor between 1.1 and $\sim$ 80 (from shorter to longer wavelengths) when the outer radius of the emitting volume is multiplied by a factor 2.0. 
Moreover, extending the emitting region to larger distances implies that the velocity of the X-ray emitting plasma (assumed to move along with the cool wind) becomes larger if the outer radius is still located inside the wind acceleration zone. This in turn increases the full width at half maximum of the lines as well as the line width at the continuum level. The effect is best seen in the red wing of the line which is most strongly depressed by the wind absorption. With a larger emitting volume, a larger fraction of the redshifted photons from the rear outer parts of the wind escape the wind since they cross layers of lower density. 
The impact of the increase of the size of the emitting region on the overall spectrum can be seen in Fig.\,\ref{fig7}.

\subsection{Inner emission boundary}

\begin{figure}
 \includegraphics[width=9cm]{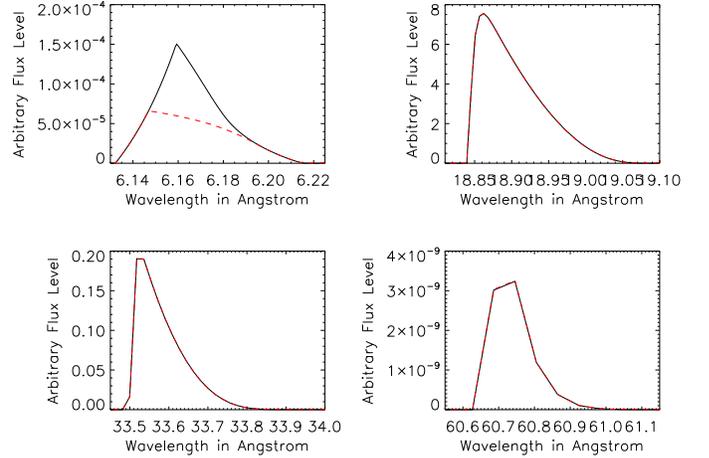}
\caption{Importance of the inner boundary of the X-ray emission region. The dashed red (respectively solid black) line profiles illustrate the results of our simulations for different lines assuming $r_{0} = 2.5$\,R$_{*}$ (resp.\ 1.5\,$R_*$) and on outer radius of the emitting region of 10$R_*$.  The huge $\tau_{*}$ value at large wavelengths (Fig.\ref{tau0}, Tab.\ref{tab1}) makes the wind opaque to X-ray photons. Consequently the X-ray photons produced in the inner part of the emitting shell are all absorbed by the cool wind material (bottom panels and right top panel).}\label{fig8}
\end{figure}

\begin{figure}
 \includegraphics[angle=90,width=9cm]{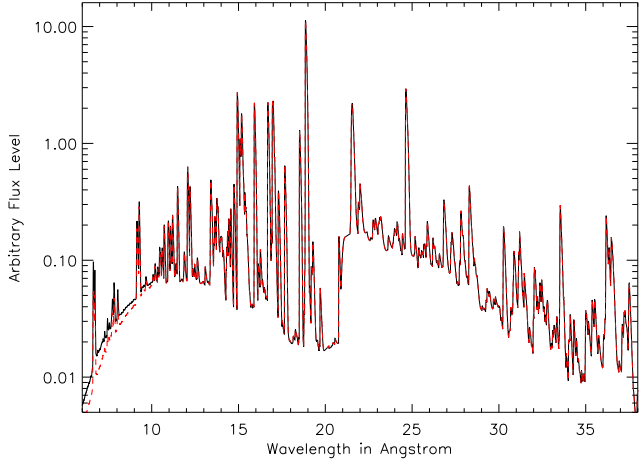}
   \caption{Same as Fig.\,\ref{fig8}, but this time for the overall X-ray spectrum.\label{fig9}}
\end{figure}

The location of the inner boundary of the emission region is one of the open questions in the understanding of the X-ray emission of massive stars. However, a variation of this parameter can have important consequences. As we pointed out in the previous section, the density of matter increases as one moves inwards in the stellar wind, and therefore the contribution to the intrinsic X-ray emission increases as well. Still, as far as the observable emission is concerned, part of this increase in emission measure is compensated by the increase of the amount of absorbing material that the photons have to cross before they escape. 
If we adopt an inner radius of 2.5\,$R_*$ instead of 1.5\,$R_*$, with an outer radius of 10.\,$R_*$, for the emitting plasma, we note that the line intensity in the short wavelength range, where $\tau_*$ is low, decreases roughly by a factor two. In the long wavelength domain, where $\tau_*$ is high, the reduction of the stellar wind absorption is more important than that of the emission. In consequence, the inner part of an X-ray emission shell is fully absorbed by the cool material. The result is that the lines at longer wavelength are rather insensitive to the location of the inner boundary of the emission region (Fig.\ref{fig8}). The same qualitative considerations hold for the overall spectrum which is shown in Fig.\,\ref{fig9}.

We note that the FIR triplet of helium-like ions provides in principle a powerful diagnostic of the inner emission boundary \citep{Leutenegger06}. However, in the case where the observed X-ray emission stems from a multi-temperature plasma with each component possibly having its own value of the inner boundary, it must be stressed that the resultant triplets are the sum (weighted by the emission measures) of the triplets in the various components. This situation renders the interpretation of the FIR diagnostics more complex than expected from a single temperature plasma.

\subsection{The contribution of the inter-clump medium}
In their work, \citet{Zsargo} have shown the need to introduce a homogeneous inter-clump medium to better fit the profile of the \ion{O}{6} $\lambda\lambda$\,1032,1038 \AA\ doublet with the model atmosphere code CMFGEN. These authors conclude that the inclusion of a small percentage $(\sim 5\%)$ of homogeneous material in between the clumps is needed to achieve a good fit of the line. Similar conclusions were reached by \citet{Sundqvist}.

\begin{figure}
\includegraphics[angle=90,width=9cm]{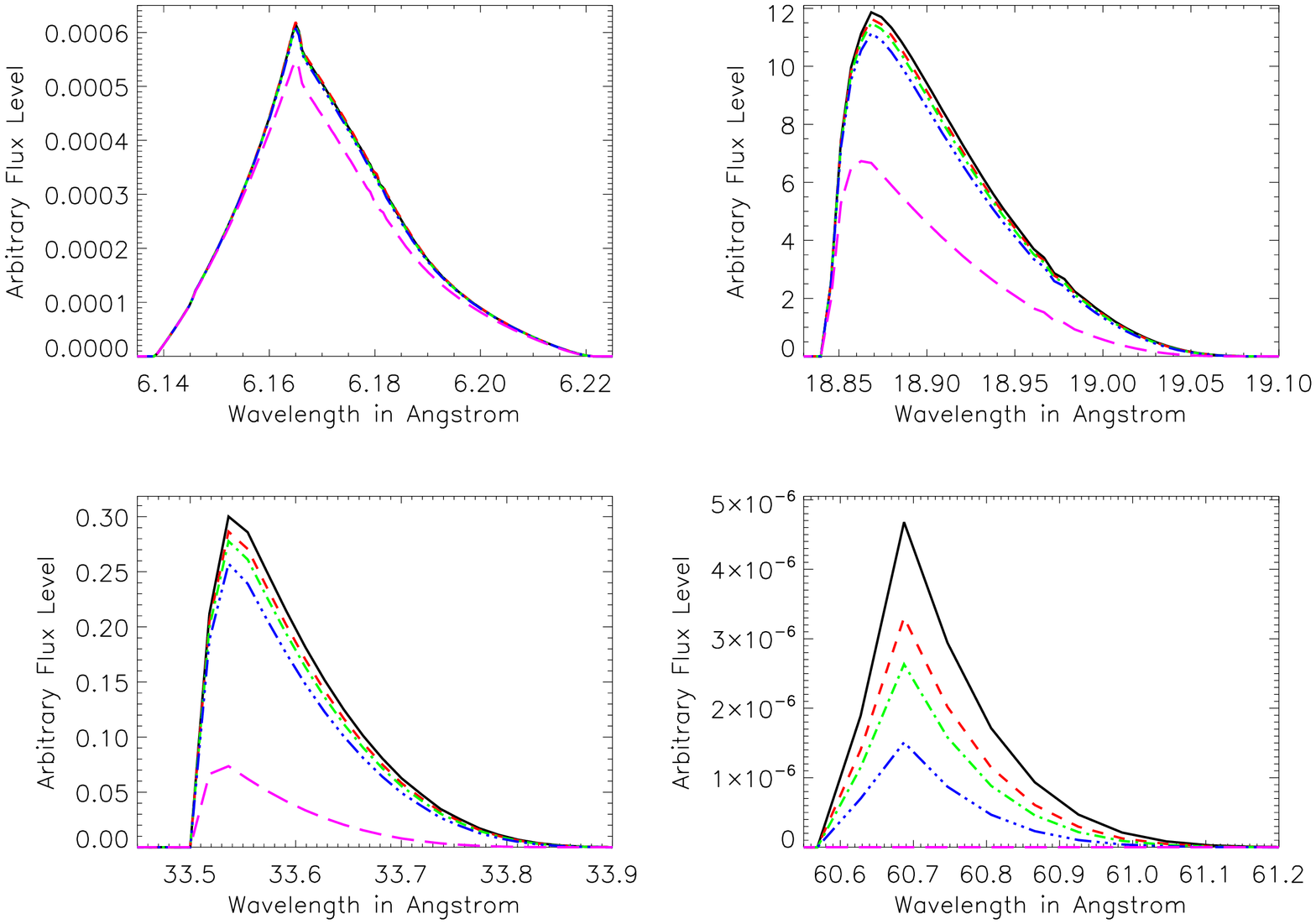}
\caption{Impact of the presence of an inter-clump medium. The solid black line profiles correspond to a fully porous wind (with $h=0.10$\,$R_*$), whilst the dashed red profiles yield the results for a fully homogeneous wind. The dot dot dashed blue, dot dashed green and long dashed magenta profiles display the situation for 3 different mixed models with respectively 90\%, 95\% and 97\% of the wind material in clumps.\label{fig10}}
\end{figure}

\begin{figure}
\includegraphics[angle=90,width=9cm]{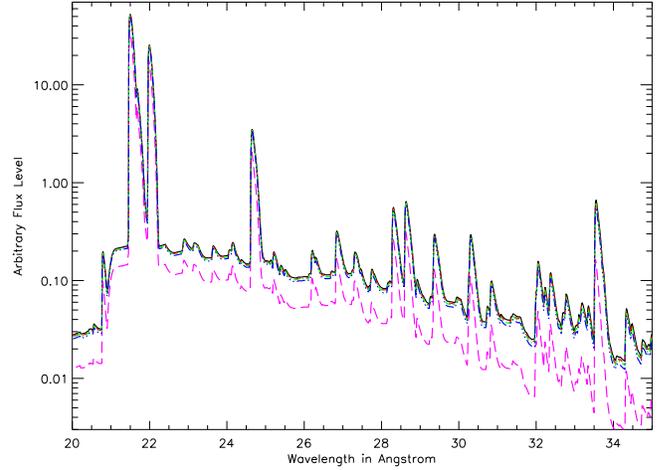}
   \caption{Same as Fig.\,\ref{fig10}, but illustrating the overall spectrum between 20 and 35\,$\AA$. \label{fig11}}
\end{figure}

\begin{figure}
\includegraphics[angle=90,width=9cm]{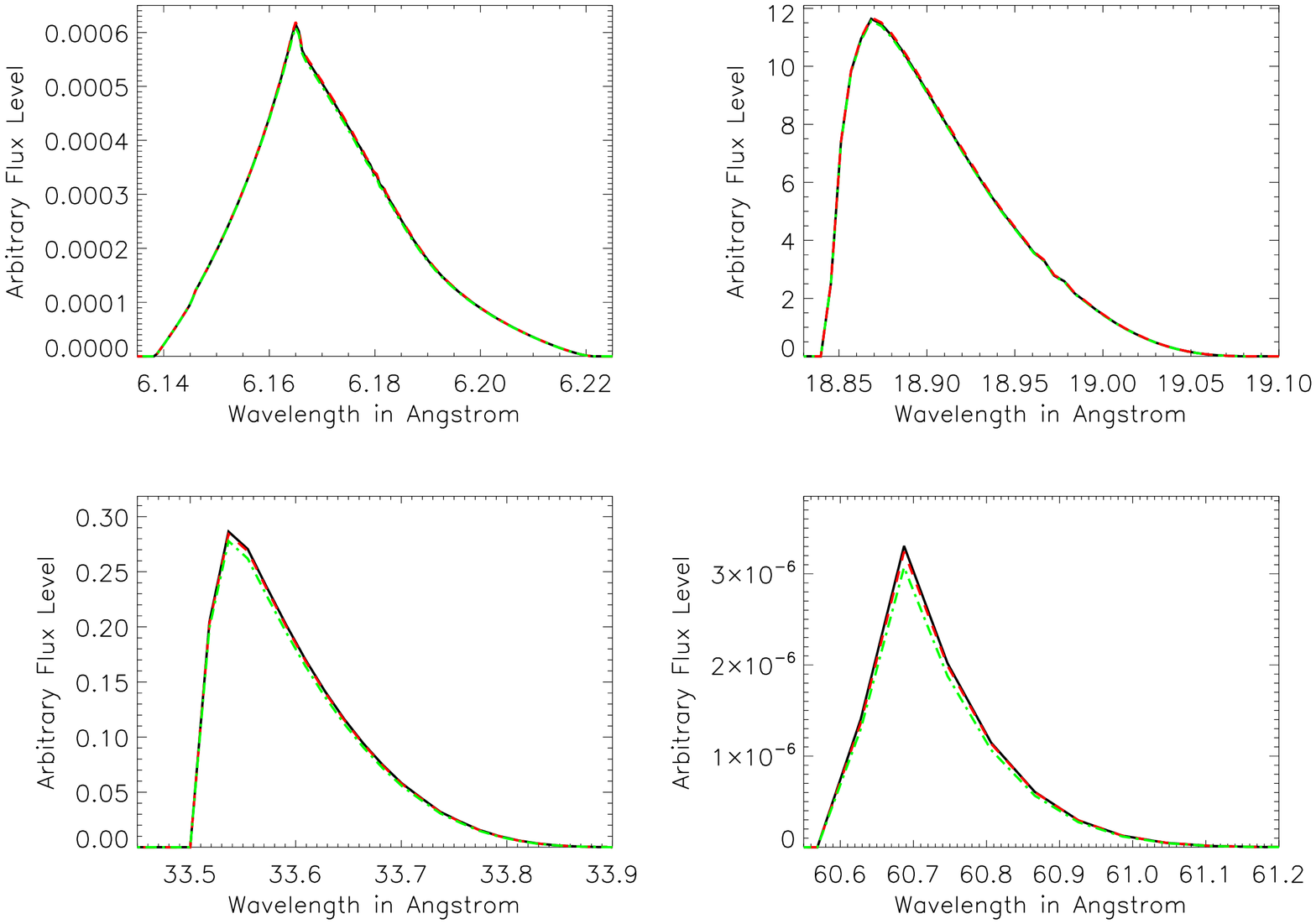}
\caption{Comparison between a mixed model with a fully clumped model (see text for details).
\label{fig12}}
\end{figure}

What is the consequence of such a tenuous, but homogeneous, inter-clump medium on the X-ray line profiles? To answer this question, we incorporate this situation in our model. In practice, we work with a constant mass-loss rate and we consider that the size of the clumps is small. We distribute an important percentage of matter (90$\%$, 95$\%$ or 97$\%$) in clumps and we compute the optical depth due to a totally clumped wind. As a first assumption, we consider that the ionization is the same in the cool clump as in the inter-clump medium\footnote{Constraining the ionization of the inter-clump medium requires hydrodynamical simulations that are beyond the scope of the present paper (see also \citealt{Sundqvist} and \citealt{Zsargo}).}. Next, we compute the optical depth for an homogeneous wind with the remaining matter. Finally  we take the sum of the two contributions and determine the absorption for a mixed wind model.  
As a homogeneous wind absorbs more photons, the intensity of the line profiles of the mixed model should actually be lower than for a fully clumped model (see Fig.\,\ref{fig10}) and the impact is stronger at longer wavelengths, and hence at larger opacities $\kappa$. These considerations are confirmed by our simulations of the overall X-ray spectrum (see Fig.\,\ref{fig11}).
\linebreak
As a second step, we  attempt to fit the spectrum of the mixed model with 97\% of matter clumped to a model spectrum for a fully clumped wind. Fig.\,\ref{fig12} illustrates the results: the individual line profiles of the mixed model are almost perfectly fitted by a fully clumped model with a slightly lower porosity parameter. This has important consequences in practice: whilst the X-ray line profiles might allow us to infer the presence of clumps and to constrain their geometry, it seems almost impossible to distinguish between a fully clumped wind and a mixed model without constraints from observations at other wavelengths (e.g.\ UV, optical).We thus conclude that X-ray line profiles alone cannot reveal the presence of an inter-clump medium.

\section{First application to the \ion{O}{8} $\lambda$ 18.97$\AA$ line of $\zeta$\,Puppis}

In the previous sections, we have discussed the impact of different parameters and of different fragmentation prescriptions on the line profiles and on the emergent X-ray spectra. 
In this section, we use our code to fit the \ion{O}{8} Lyman $\alpha$ doublet $\lambda$18.97 $\AA$ and the neighbouring \ion{N}{7} $\lambda$ 19.36 $\AA$, $\lambda$ 19.82 $\AA$ lines of the massive star $\zeta$\,Puppis and determine some stellar wind parameters. We decided to work on an extended region around the strong \ion{O}{8} line since Fig.\,\ref{figpre2} reveals the presence of weak lines that affect the wings of the \ion{O}{8} line and produce a pseudo-continuum that contributes of order 5\% of the flux of the \ion{O}{8} line. For this purpose, we have extracted, reduced and combined 18 RGS spectra of $\zeta$\,Puppis from the $XMM-Newton$ archive (\citealt{Naze}; the different aspects of the data reduction are treated in detail in that paper).  
We also apply the absorption by the interstellar medium ($N_{H}=8.9\times10^{19}\mbox{cm}^{-2}$, \citealt{DS}) to the synthetic spectra. Then, we convolve the spectra with a Lorentzian profile with a half-width at half-maximum of 0.07 $\AA$ to account for the instrumental resolution of the RGS spectrograph onboard  $XMM-Newton$. Finally, we compare the results to the observation.

\begin{figure}
\includegraphics[angle=90,width=9cm]{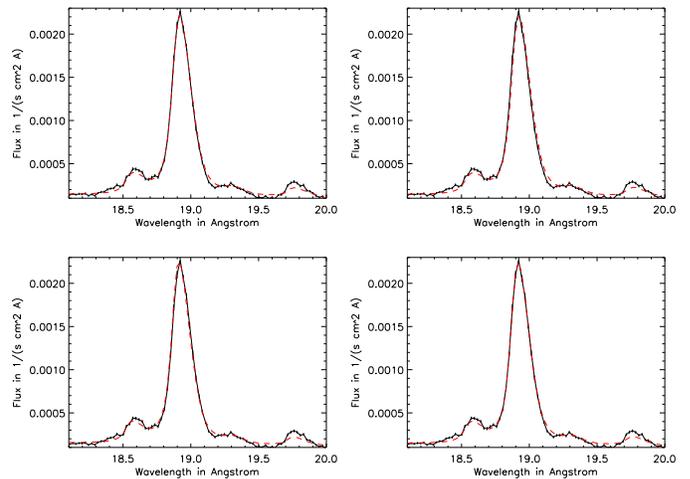}
\caption{Fit of the spectral region around the \ion{O}{8} $\lambda$18.97 line of $\zeta$\,Puppis. There exists a degeneracy on the determination of the mass-loss rate, the porosity of the wind and the hot gas filling factor. {\it Top panels}: Profiles derived for a homogeneous wind (on the left, model A in dashed red line) and a porous wind with the same mass-loss rate (on the right, model B in dashed red line) compared to the observational data (in solid black line). {\it Bottom panels}: same but for a different mass-loss rate (model C on the left, model D on the right). Model parameters are given in Table 2. }
\label{fig13}
\end{figure}

\begin{figure}
\includegraphics[angle=90,width=9cm]{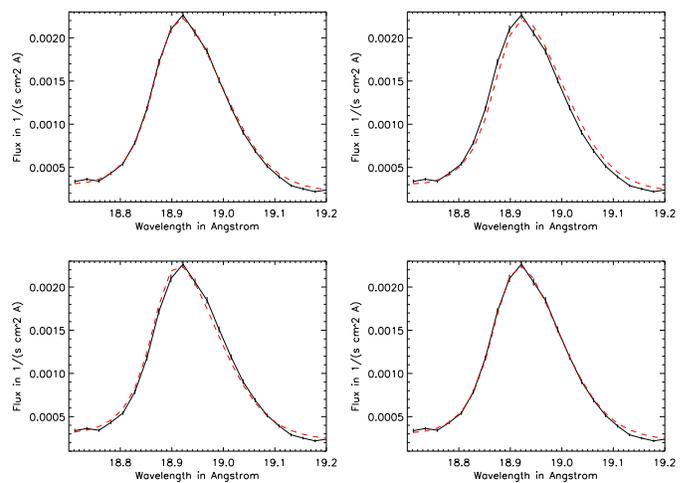}
\caption{Same as Fig.\,\ref{fig13} but zooming in on the \ion{O}{8} $\lambda$18.97 line.}
\label{oviiizoom}
\end{figure}

As shown in Fig.\,\ref{fig13} and \ref{oviiizoom}, we obtain a degeneracy of  the stellar wind parameters. Indeed, for the \ion{O}{8} line profile, we demonstrate that we can use a homogeneous or a porous wind model to fit the data equally well (see Fig.\,\ref{fig13} and Table 2) though with different values for the hot gas filling factors. Note that these values are in agreement with the results of \citet{Hillier93}. We only need to adapt the mass-loss rate and the abundance of oxygen to obtain similar results. 
In a denser wind (i.e.\ with a larger mass loss rate), the increase of absorption dominates the increase of X-ray photon production at longer wavelengths where the opacity is largest. In consequence we have two ways to increase the number of X-ray photons which escape from a denser wind. The first solution is to increase the number of X-rays resulting from shocks. To first order, we could imagine that the number of shocks, and hence the hot gas filling factor, would be larger in a denser wind (Tab.\,\ref{tab2}, models A and C). Alternatively, we can increase the number of photon which escape. Consequently, we have to increase the free mean path of photons in the stellar wind. This can be achieved with a more porous wind (Tab.\,\ref{tab2}, models A and D). Normally, the line profiles of the two previous solutions should be different. Unfortunately, the instrument resolution is not sufficient to distinguish the subtle differences.

\begin{figure}
\includegraphics[angle=90,width=9cm]{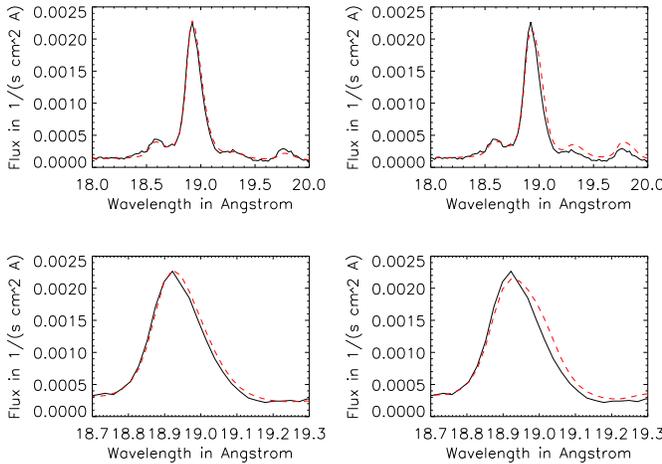}

\caption{ {\it Left}: Comparison  of the \ion{O}{8} $\lambda$18.97 $\AA$ line profile of the non-void inter-clump medium model E (in dashed red line) to the observational data (in solid black line).
{\it Right}: Comparison of the anisotropic model F (in dashed red line) to the observational data (in solid black line). Model parameters are given in Table 2. 
\label{fig14}}
\end{figure}

Furthermore, we can also fit this line equally well with a model where the inter-clump medium is not void (Fig.\,\ref{fig14}, on the left).

It is thus obvious that a single line does not permit to discriminate between models. These preliminary results therefore emphasize the necessity to work on the full X-ray spectrum. The analysis of the full RGS spectrum is beyond the scope of the present paper, but will be presented in a future work (Herv\'e et al., in preparation). As we demonstrated in Sect.\ 3, each parameter has a different impact over different parts of the spectrum and  we are therefore confident that some degeneracies on e.g.\ porosity or mass-loss rate, will then be lifted. 
 
\begin{figure}

\includegraphics[angle=90,width=9cm]{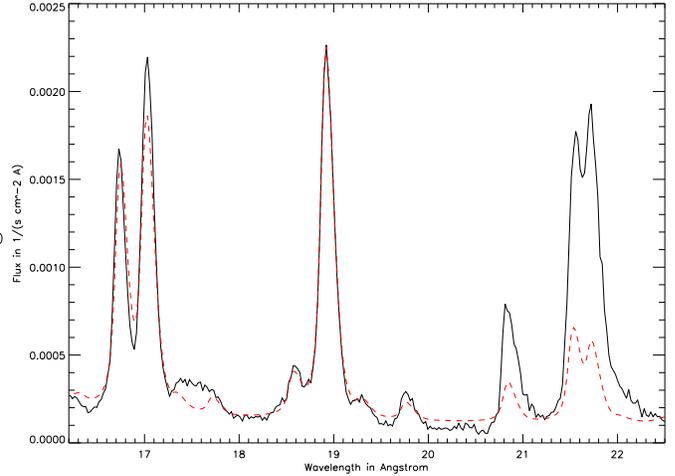}
\caption{Even though the \ion{O}{8} $\lambda$18.97 $\AA$ line is well fitted with an homogeneous model (Model A in dashed red line), the \ion{O}{7} triplet and \ion{N}{7} lines in the model are too weak. That indicates the need to use a second temperature in order to fit the whole spectrum of $\zeta$\,Puppis. Consequently, the hot gas filling factor of the first plasma temperature will be lower in order to take account of the contribution, in the \ion{O}{8} intensity and line profile, of the second plasma temperature emission. 
\label{fig15}}
\end{figure}

Nevertheless, already on this small region of the spectrum, three results are found. First, the possibility of an anisotropy (i.e., $\mu \neq 1 $, in the fragmentation prescription) of the wind absorption appears unlikely. Indeed, the flux from the back side of the star would be too large (Fig.\,\ref{fig3}, on the top right) and hence the width of the synthetic line emission would be too large in comparison to the observations (Fig.\,\ref{fig14}, on the right). Second, a good fit to the \ion{O}{8} line and its close neighbours is only found if one uses an overabundance of N and a depletion of O (Tab.\,\ref{tab2} and Fig.\,\ref{fig13}). Third, a quick look at a wider wavelength band (Fig.\ref{fig15}) shows that a single temperature plasma cannot reproduce the whole spectrum of $\zeta$\,Puppis. A multi-temperature plasma model is required for a full analysis of the X-ray spectrum. Thus the observed line profiles and intensities are a combination of the contribution of the different temperature plasmas present in the wind. 

Finally, it should be noted that part of the degeneracy of the results is due to the limited resolution of present-day high-resolution X-ray spectrographs. In Sect.\,3, we show the impact on a line profile of a change in porosity length or inner radius for example, but the convolution by the response matrix of current instruments washes out part of the difference induced by the different parameters.

\section{Conclusions}

In this paper, we have studied the impact of the two different prescriptions of stellar wind fragmentation on the emergent broad-band X-ray spectrum. For spherical clumps with an isotropic opacity, the differences between the two prescriptions are the most obvious when comparing the overall X-ray spectrum. Indeed, in this case, we find that a given simulation with a single value of the fragmentation frequency cannot be represented by a spectrum calculated with a single value of the porosity parameter.
However, in practice, with the quality of current high-resolution X-ray spectra of O-type stars, it seems almost impossible to distinguish the two models in a real case when working on a line-by-line basis. This is because the differences are often small and are washed out by the coarse resolution of present-day instruments. Moreover, the differences are most easily seen when comparing the shape of the full spectrum which is seldom done. In addition, one must keep in mind that the shape of spectrum is also affected by the amount of interstellar absorption and the fact that the X-ray emission from a real stellar wind most likely spans a range of plasma temperatures rather than a single temperature as assumed here (see e.g.\ \citealt{zhekov07}; \citealt{2XMM}). In both prescriptions we find that the X-ray emission is more intense in a fragmented stellar wind than in a homogeneous wind.\linebreak
The introduction of an anisotropic opacity in the fragmentation frequency prescription yields spectra that are more intense than in the isotropic case and of course than in an homogeneous wind. Sophisticated, 3-dimensional hydrodynamical calculations might eventually be needed to clarify the issue of the geometry of the clumps.

As pointed out previously, both models rely on a series of assumptions and, currently, we do not have enough knowledge on the stellar wind fragmentation to privilege one model over the other.\linebreak
Our investigation of the impact of the various model parameters on the emerging spectrum showed that both the position and the size of the X-ray emitting plasma are very important. An increase in the size leads to a more luminous spectrum, although we have shown that the amplitude of the intensity increase depends upon wavelength.

For a given mass-loss rate, the differences between fully clumped and homogeneous wind models are quite large. However, in light of the simulations presented here, it will be very difficult to estimate correctly the value of the porosity if we include the effect of an inter-clump medium. In fact, the incorporation of a few percent of the wind material in the inter-clump medium yields the same emergent spectrum as obtained for a purely clumped model with a slightly less fragmented wind.\linebreak

Finally, we present preliminary fits of the region near the \ion{O}{8}  $\lambda$ 18.97$\AA$ emission line of $\zeta$\,Puppis with our code. Our first results reveal that anisotropic clumps are unlikely, and that non-solar abundances (overabundance of N, depletion of O) are needed to reproduce the X-ray spectrum. However, there is also a degeneracy of stellar wind parameters.  Even though a porous wind is closest to the observational data, we can not totally exclude an homogeneous wind. The mass-loss rate, the porosity parameter and the hot gas filling factor are strongly tied. However, simultaneous fits of the total spectrum will likely allow to reduce the degeneracy on the porosity and other parameters. But, it appears that a single temperature X-ray plasma is insufficient to fit the global spectrum of $\zeta$\,Puppis. This result implies the necessity to use several temperatures and we will have to include the contribution of each temperature plasma to the line intensity and profile of \ion{O}{8}  $\lambda$ 18.97$\AA$ (and all the other lines) but also to the continuum.

\begin{table}[htbp]
\caption{Lines considered in this study and the wind optical depth extracted from Fig.\,\ref{tau0} and corresponding to our stellar and wind parameters}
\label{tab1}

\begin{center}
\begin{tabular}{llll}

\hline
\hline
Line    & Wavelength ($\AA$)   & $\tau_{*}$ \\
\hline
\ion{Si}{14}  & 6.18                 & 1.8         \\
\ion{O}{8}    & 18.97                & 21.9        \\
\ion{C}{6}    & 33.73                & 46.2        \\
\ion{Si}{8}  & 61.02                & 187         \\

\hline
\end{tabular}
\end{center}
\begin{list}{}{}
\item[$^1$] We are aware that the \ion{Si}{8} $\lambda$ 61.02 line will be impossible to observe in practice because it is expected to be intrinsically faint and most of all because of the huge absorption by the interstellar medium at these wavelengths. However, here we use this line to illustrate the behaviour of lines at high $\tau_{*}$ values.
\end{list}
\end{table}

$  $

\clearpage

\begin{table*}[htbp]
\caption{Stellar and wind parameters for the fit of the \ion{O}{8} $\lambda$18.97 $\AA$ line in the RGS spectrum of $\zeta$\,Puppis with $kT_{plasma}=0.217$ keV, $R_{*}=18.6$ R$_{\sun}$, $R_{inner}=1.5 R_{*}$, $R_{outer}=10 R_{*}$, $v_{\infty}=2250$\,km\,s$^{-1} $, $\frac{O_{*}}{O_{\sun}}=0.3$ (by number) and $\frac{N_{*}}{N_{\sun}}=4.0$ (by number).}
\label{tab2}

\begin{center}
\begin{tabular}{cccccc}

\hline
\hline
Model & $\dot{M}$ & $f_{hot~gas}$ & $h$ & note & $\chi^{2}_{\nu}$\\
        &  ($10^{-6}$\,M$_{\sun}$\,yr$^{-1}$) &  & ($R_*$)\\
\hline
A & 3.0 &  0.033  & 0.00  & & 4.10\\
B & 3.0 &  0.028  & 0.10  & & 5.44\\
C & 4.0 &  0.036  & 0.00  & & 4.39\\
D & 4.0 &  0.029  & 0.10  & & 3.92 \\
E & 4.0 &  0.028  & 0.10  & 5$\%$ of inter-clump & 5.31\\
  &      &         &       &  matter &\\
F & 4.0 &  0.026  & n$_{0}=1.7 10^{-4}$\,s$^{-1}$ & $\mu \neq 1$ (fragmentation & 8.32\\
  &      &         &                      &  prescription)\\
\hline
\end{tabular}
\end{center}

\end{table*}

\begin{acknowledgements}
This research is supported by the FNRS (Belgium) and by the Communaut\'e Fran\c caise de Belgique - Action de recherche concert\'ee (ARC) - Acad\'emie Wallonie--Europe and an XMM+INTEGRAL PRODEX contract.
(Belspo)
\end{acknowledgements}

\end{document}